\documentclass[aps,prl,twocolumn,groupedaddress,floatfix]{revtex4}

\usepackage{graphicx}

\graphicspath{{FiguresAcoustics/}} 

\begin{document}


\title{Low Frequency Acoustic Resonance Studies of the Liquid-Vapor Transition in Silica Aerogel}

\author{Tobias Herman}
\email[]{therman@phys.ualberta.ca}

\author{John R. Beamish}
\email[]{beamish@phys.ualberta.ca}

\affiliation{Department of Physics, University of Alberta,
Edmonton, Alberta, Canada T6G 2J1}


\date{\today}

\begin{abstract}
Fluid phase transitions in porous media are a powerful probe of
the effect of confinement and disorder on phase transitions.
Aerogel may provide a model system in which to study the effect of
dilute impurities on a variety of phase transitions.  In this
paper we present a series of low frequency acoustic experiments on
the effect of aerogel on the liquid-vapor phase transition.
Acoustic resonators were used to study the liquid-vapor transition
in two fluids (helium and neon) and in two different porosity
aerogels ($\sim 95\%$ and $\sim 98\%$).  While effective
coexistence curves could be mapped out, the transition was
sometimes difficult to pinpoint, leading to doubt as to whether
this transition can be treated as an equilibrium macroscopic phase
transition at all.
\end{abstract}


\maketitle

\section{Introduction}

Fluids confined in porous media provide a valuable system in which
to study the effects of disorder and confinement on phase
transitions and critical behavior.  Helium, with its rich phase
diagram, has proven a particularly important testing ground.  For
instance, when liquid $^4$He is confined in porous Vycor glass,
the superfluid density varies with the same exponent as near the
bulk lambda transition, but the heat capacity anomaly is greatly
suppressed\cite{Zassenhaus99}.  Films of $^4$He adsorbed on Vycor,
on the other hand, share some features with the
Kosterlitz-Thouless transition in 2-dimensional films (e.g.
superfluid transition temperatures decrease with film thickness)
but have significant heat capacity peaks
\cite{Murphy90,Crowell97}.

Silica aerogels, with their tenuous structure and extraordinarily
low densities, provide a unique opportunity to introduce small
concentrations of impurities into helium in a controllable way.
Experiments include a variety of transitions which are affected
differently by the presence of aerogel.   For example, the lambda
transition for $^4$He remains very sharp\cite{Chan88,Yoon98} but
with a non-bulk critical exponent for superfluid density, while
for $^3$He, aerogels suppress or even completely eliminate
superfluidity\cite{Matsumoto97,Porto99}, resulting in a zero
temperature ``quantum phase transition.''   A major difference
between these superfluids is their correlation length.  In $^4$He
it is very small and only approaches the characteristic length
scale of the aerogel structure at temperatures very close to the
lambda point; in $^3$He the correlation length is much larger,
comparable to that of the aerogel at all temperatures.  The
aerogel strands also couple more strongly to some order
parameters, e.g. to the concentration in a binary fluid mixture,
so we would expect a transition such as phase separation to be
strongly affected by an aerogel.   For $^3$He-$^4$He mixtures, the
presence of 2\% or even 0.5\% silica (i.e. aerogels with porosity
of 98\% or 99.5\%) has dramatic
effects\cite{Chan96,Kim93,Mulders95} on the entire phase diagram,
causing the phase separation curve to detach from the lambda line
and stabilizing a region of dilute $^4$He superfluid inaccessible
in bulk mixtures.

The liquid-vapor transition has an order parameter (density) which
couples directly to a porous medium via the Van der Waals
interaction. It is well known that in small pores the attraction
to pore surfaces stabilizes the liquid phase at pressures well
below bulk saturation. In most porous media the variety of
adsorption environments leads to a gradual transition in fluid
density from a ``pore vapor'' to a ``pore liquid'' phase, rather
than the sharp transition of bulk fluids. Adsorption isotherms are
used to study this capillary condensation, which is often
accompanied by hysteresis between adsorption and desorption,
indicating the importance of metastable liquid-vapor
configurations.  The traditional picture of capillary condensation
involves a liquid covering the surface of a simple pore (e.g. a
cylinder) and a liquid-vapor interface with a negative curvature
needed for surface tension to stabilize the liquid phase. However,
this picture breaks down in aerogels and near the liquid-vapor
critical point (LVCP).

Aerogels are better described as a dilute network of
interconnected strands than as a collection of closed pores and so
we might expect a liquid film to have a positive rather than
negative curvature.  Aerogel strands are very narrow (diameters of
a few nm) and close to the LVCP thermal fluctuations grow from
atomic to macroscopic scales.  In this regime, many thermodynamic
properties diverge (e.g. compressibility) or tend to zero (e.g.
surface tension) and the thickness of the liquid-vapor interface
itself exceeds the strand diameter or even the effective pore
size.  It is then necessary to picture the aerogel as a random
perturbation affecting the fluid phases in some averaged way.

In bulk fluids, the liquid-vapor transition falls into the Ising
universality class and deGennes\cite{Brochard83,deGennes84} has
suggested that fluids in porous media may provide a realization of
the random field Ising model. Studies of binary fluid
separation\cite{Goh87,Monette92,Lin94,Frisken95a,Valiullin02} in
porous media lend some support to this theory, although in some
cases the behavior can also be described within a single-pore
model which neglects the presence of disorder in the medium.
Experimental work on near-critical single-component fluids
includes xenon and SF$_6$ in dense porous
glasses\cite{Machin99,Thommes94,Thommes95}. These studies showed a
slight narrowing in the phase separation curve, with the vapor
branch shifted to higher densities than in the bulk fluid. In
addition, the termination of this curve, sometimes referred to as
the ``capillary critical point,'' was shifted to a temperature
below the bulk LVCP. Despite the relatively well defined pore
geometries of these systems there is no general agreement about
how to picture the behavior near the LVCP.

The first liquid-vapor experiments involving
aerogels\cite{Wong90-2567} showed a reduced critical temperature
for $^4$He in a 95\% porosity sample and a coexistence region
which was dramatically narrowed (by an order of magnitude).  Below
the critical point, isotherms appeared to have discontinuous
density jumps between vapor- and liquid-like phases and no
hysteresis was observed, consistent with an equilibrium
liquid-vapor transition. The surprisingly narrow coexistence
curve, determined from heat capacity and isotherm measurements,
could be fit using the bulk critical exponent. Light scattering
measurements with N$_2$ in aerogel\cite{Wong93} also showed a
coexistence curve that was narrower than for bulk, but
substantially broader than for $^4$He.

More recently, a mechanical pendulum\cite{Gabay00-99,Gabay00-585}
was used to study $^4$He in a similar aerogel. Thermal response
times were very long but the isotherms showed hysteresis between
filling and emptying, even near the critical point, and appeared
to have finite slopes at all temperatures. This behavior is
characteristic of capillary condensation, rather than equilibrium
coexistence and critical behavior.   Preliminary ultrasonic and
acoustic resonator measurements in our
laboratory\cite{Tan00,Herman02,Herman03} showed hysteresis for
both neon and helium in aerogels and implied phase separation
curves for which were substantially broader than the $^4$He curve
in Ref.~\onlinecite{Wong90-2567}.  A recent direct measurement of
density fluctuations in carbon dioxide confined in
aerogel\cite{Melnichenko04} showed little evidence for a diverging
correlation length at the LVCP, suggesting that the disorder
introduced by aerogel structure may limit the range of thermal
fluctuations.

In parallel with this acoustic study of fluids in aerogel we have
performed
measurements\cite{Beamish03-340,Beamish04-339,tobyPhD,Herman05a}
of helium density in aerogels along adsorption and desorption
isotherms at temperatures ranging from far below to well above the
bulk LVCP. While the aerogels filled and emptied over narrow
pressure ranges, there was no direct evidence for a discontinuous
change in fluid density along any of the isotherms.  Hysteresis
between the adsorption and desorption branches was evident until
very close to the critical point in both 95\% and 98\% porosity
aerogels, in contrast to Ref.~\cite{Wong90-2567}. Furthermore,
these experiments provided a direct measurement of the thermal
relaxation within the aerogel
--- even the very thin samples used ($\sim0.5$mm thick) exhibited
equilibration times of up to several hours.

The differences between the behavior seen in the early $^4$He heat
capacity experiments and that in more recent isotherm measurements
are puzzling, but could reflect differences in how the
measurements were made.  The heat capacity and light scattering
measurements followed isochores, in cells almost completely filled
with aerogel, so only small amounts of helium were adsorbed or
desorbed when the temperature was changed. Subsequent isotherm
measurements were made in cells which contained large amounts of
bulk fluid and involved substantial changes in the adsorbed helium
density.   Although the different measurements used aerogels of
similar porosity (about 95\%), it is also possible that the
hysteresis and the width of the liquid-vapor coexistence curve are
sensitive to small variations in aerogel density or structure.
Simulations of fluids in aerogels\cite{Krakoviack01} indicate that
the liquid-vapor behavior is also controlled by the relative
strengths of adsorbate-adsorbate and adsorbate-substrate
interactions, so variations in the width of the coexistence curves
for helium and nitrogen could reflect the interactions in these
fluids.

The experiments described in this paper were designed with several
goals in mind.  We wanted to measure a fundamental thermodynamic
parameter, to complement the information available from adsorption
isotherms and provide a more complete picture of critical behavior
of fluids in aerogels.   We looked for a probe that, as well as
being sensitive to phase separation, could provide thermodynamic
information in the single phase region above the critical
temperature where measurements are not complicated by the presence
of a liquid-vapor interface.  In order to avoid large amounts of
fluid moving in and out of the aerogel and to allow direct
comparison to the earlier heat capacity measurements, we tried to
minimize the amount of bulk fluid surrounding the aerogels and
took data along isochores.  We wanted to measure coexistence
curves in more detail and over a broader temperature/density range
around the LVCP than previous experiments.  By using different
fluids in the same aerogel we could study how the behavior varied
with interaction strength and by comparing the behavior in
aerogels with different porosities we could see how it depended on
impurity density.

To meet these goals, we built acoustic
resonators\cite{Herman02,Herman03} and used them to study helium
and neon in aerogels with porosities near 95\% and 98\%. Acoustic
resonators have proven their value in the study of superfluidity
in aerogels, for example by measuring the speed of sound in $^3$He
confined in aerogels\cite{Golov99,Lawes03}.  A resonator measures
compressibility, one of the fundamental thermodynamic properties
that characterize critical behavior.  The isothermal
compressibility of a fluid diverges at T$_c$ but low frequency
acoustic measurements probe the adiabatic compressibility, which
is dominated near T$_c$ by the heat capacity's behavior allowing,
for example, the heat capacity exponent $\alpha$ to be determined
for bulk fluids\cite{Thoen74-1311}.

Acoustic resonances probe the entire sample and their frequencies
are sensitive to the large scale inhomogeneities associated with
phase separation.   When aerogels are present, the damping of
resonant modes provides additional information about dissipation
due to processes such as microscopic heat flow and redistribution
of fluid in response to adiabatic compressions.  Aerogel
complicates matters as well, since most resonator modes then have
both compressional and shear components.  The resonant frequencies
also depend on boundary conditions, so even a relatively small
bulk-fluid volume around the aerogel can have a significant
effect. However, even if it is not possible to extract
quantitative values of fluid compressibility, both the resonance
frequencies and their amplitudes can be used as markers of phase
separation.  Since sound modes depend directly on density, which
for high porosity aerogels is dominated by the fluid, acoustic
resonators provide a sensitive method by which to map out the
coexistence curve, as has been done for bulk
fluids\cite{Colgate90-198}.

In this paper we present detailed results of our acoustic
resonator studies of liquid-vapor behavior in aerogels.  We made
measurements with helium and neon in aerogels similar to those
used in earlier experiments ($\sim$95\% porosity) and with helium
in a lower density ($\sim$98\% porosity) aerogel.
 Resonator measurements with bulk fluid were easily interpreted, but
were much more complicated when the aerogels were present, with
unexpected splitting of resonances and apparent mode crossings.
Nonetheless, we were able to map out coexistence curves for the
three systems described above.  The results for helium were
consistent with previous isotherm
measurements\cite{Gabay00-585,Herman05a} but in neither aerogel
was there evidence for coexistence curves as narrow as those
reported from heat capacity measurements\cite{Wong90-2567}. The
helium and neon coexistence curves in the 95\% porosity aerogel
had similar widths, both somewhat narrower than in bulk. The
helium coexistence curve in the 98\% porosity aerogel was wider
than in 95\% aerogel, although still narrower than for bulk,
consistent with the idea that a lower density aerogel has less
effect on the liquid-vapor transition.

\section{Experimental Details}
Our cells were cylindrical resonance chambers capped by thin
membranes; piezoelectric transducers mounted on the two membranes
acted as speaker and microphone respectively. When filled with a
single phase fluid (without aerogel) the resonance frequencies
($f_n$) were those of a closed pipe:
\begin{equation}f_n = n \left( \frac{v}{2L}\right)\end{equation}where
\textit{v} is the speed of sound in the fluid, \textit{L} the
length of the cavity, and \emph{n} is any positive integer. Due to
the low density and compliant nature of aerogel, the resonant
frequencies in the presence of aerogel were similar to those of
the bulk fluid resonator. To extract the actual fluid
compressibility from this data requires a detailed analysis which
takes into account both the solid backbone of the porous medium
and the fluid within the medium. Analysis tools were pioneered by
Biot, and expanded by others\cite{Biot56a,Biot56b,Forest98}, but
experimental factors complicated our data, preventing us from
extracting the critical behavior of the compressibility. Hence,
the features of the resonant frequency along isochores were used
solely to map out the coexistence curve.

The two aerogel densities studied were $110 \frac{kg}{m^3}$ and
$51 \frac{kg}{m^3}$, corresponding to porosities of 95\% and
slightly less than 98\% respectively.  They are referred to as
aerogels ``B110'' and ``B51'' throughout this paper.  The aerogel
samples were synthesized in our lab from tetramethyl orthosilicate
(TMOS) using a standard one-step base catalyzed method followed by
supercritical extraction of the methanol solvent\cite{Poelz82}.
The samples were approximately 2cm long cylinders, 1.2cm in
diameter.

\begin{figure}
\includegraphics{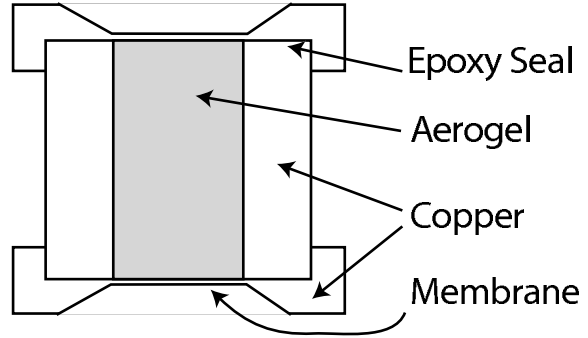}
\caption{\label{HeCell}A schematic of a cell used to investigate
helium in silica aerogels.  The neon cell was similar, but
required an indium seal rather than an epoxy seal, due to the
higher pressures.}
\end{figure}

The copper cells were capped by $\sim\frac{1}{3}$mm copper
membranes and the cavities were machined to fit the precise
dimensions of the individual aerogels used, to reduce bulk fluid
in the cell. Each sample was about 2cm long; the cavity was
machined about 0.5mm shorter than the aerogel to ensure good
contact with the membranes once assembled. Slight irregularities
in the aerogel sample resulted in up to 2\% of the resonator
cavity being filled with bulk liquid. Figure~\ref{HeCell}
shows a cell used for helium. The caps on the helium resonators
were sealed and held in place using epoxy\footnote{Bipax Tra-Bond
(BA-2151) epoxy}; the pressure was kept below 0.5MPa to avoid cell
damage.  The neon resonator caps were sealed with indium and held
in place by bolts; the membranes were slightly thicker and could
withstand pressures up to 5MPa.

A piezoelectric transducer (0.25mm thick PZT disc with silver
electrodes) was soldered onto the outside face of each acoustic
membrane and electrical contact to the top face of the transducer
was made by directly soldering a thin copper wire to the
transducer. The intrinsic resonance of these transducers was in
the MHz range, far from interfering with the fluid resonances
which were in the kHz range. The sinusoidal excitation frequency
was swept over a range that included the fundamental resonant
frequency of the system and usually the first harmonic frequency
while the response was measured by a lock-in amplifier.
Occasionally other resonances interfered with the signal (e.g.\
mechanical resonances of the membranes themselves), especially at
higher frequencies ($> 5$kHz).

The neon resonator was mounted on a closed cycle refrigerator,
suspended from  braided copper cables to adsorb vibrations while
providing a thermal link to the baseplate of the cryostat. Its
temperature was controlled using a 100$\Omega$ platinum resistance
thermometer and a 200$\Omega$ thick film resistor, both mounted on
the cell body. The helium resonators were mounted below the 1K pot
of a helium cryostat. Its temperature was controlled using a
calibrated germanium resistive thermometer and 200$\Omega$ thick
film heater. The B110 helium resonator was cooled by admitting a
few mbar of exchange gas into the vacuum can, while the B51 helium
resonator was cooled through a weak thermal link to the 1K pot.

All thermometers used in this work came with manufacturer's
calibrations.  However, the platinum resistance thermometer used
for the neon experiments showed signs of a deviation from its
calibration.  To address this, the bulk neon coexistence curve was
mapped out and the critical temperature was used as a fixed point
to adjust the thermometer readings. The recalibration used a
coaxial capacitor with neon as the dielectric between capacitor
plates; it indicated an error in thermometer calibration of about
$-120$mK. Some tests of the resonator technique using bulk neon
indicated a comparable magnitude for the thermometer error.
Combined with the capacitive measurements above these indicated
that the thermometer calibration had shifted by more than 100mK
and that it seemed to be slowly drifting over time, thus limiting
the accuracy with which we could determine temperatures along the
neon coexistence curve. The germanium thermometer used for the
helium resonators was checked against the liquid-vapor critical
point of $^4$He and found to agree within 3mK.

The density of neon in our experiment was calibrated by dosing the
cell with neon from a gas handling system of known volume.  Helium
density was calibrated using a direct capacitive density
measurement of helium in aerogel slices of the same density as
used in the resonators.  The method used for those measurements is
described elsewhere\cite{tobyPhD,Herman05a}.

Data collection was fully automated; once the cell reached a set
temperature point, it remained at that temperature for a
prescribed ``equilibration time'' while the aerogel's internal
temperature settled.  A full acoustic spectrum was then collected
(such as that shown in Fig.~\ref{BulkNeonSpectrum}, and the system
moved onto the next temperature point. The resonant frequency was
also tracked and logged during equilibration, although there were
problems with spurious resonances and irregular peak shapes. The
log files were valuable in showing how quickly the cell
equilibrated, but the final data were taken from the full spectrum
data files.

Filling and emptying of the sample between isochores were
performed manually. During filling and emptying the cell was held
well above the critical point to eliminate problems with latent
heat or the equilibration of any liquid-vapor interfaces in the
cell.  Also, at no point during data collection was the cell
allowed to cool below 43.500K for the neon in B110, 4.800K for
helium in B110, or 5.000K for helium in B51. By keeping the
temperatures in the cell close to the critical temperature we
avoided possible damage to the aerogel from surface tension
induced strains.  In another study we have measured the
deformation of these aerogels during adsorption and desorption of
low surface tension fluids\cite{tobyPhD,Herman05b} and found that
even liquid helium can cause significant deformation for gels with
porosities of 95\% and greater.

\section{Results}
Each ``coexistence curve'' described in this paper was constructed
by locating the transition temperature along each of a series of
isochores, constructed in turn from a series of acoustic spectra
taken as the temperature was varied. A phase change appeared as a
sharp feature in the isochore --- usually a dip or a kink in the
resonant frequency, often accompanied by a sharp feature in the
resonance amplitude. It will be shown that although the form of
the feature was well defined in bulk fluid experiments, it was not
as simple to locate the transition in the aerogels.

\subsection{Bulk Neon}

\begin{figure}
\includegraphics[width=\linewidth]{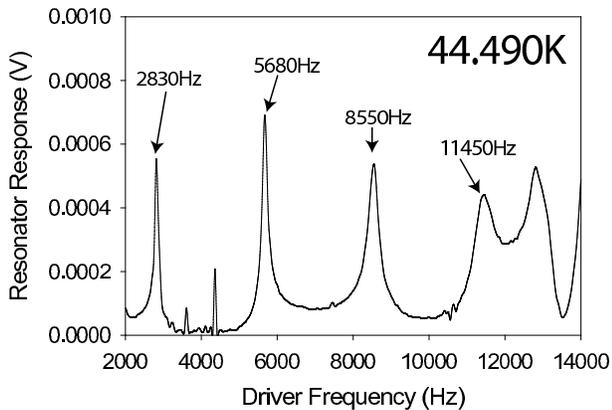}
\caption{\label{BulkNeonSpectrum}Bulk neon acoustic spectrum near
its critical point. This isochore corresponds to $\rho \approx 475
\frac{kg}{m^3}$ and T=44.490K, just above the critical
temperature.}
\end{figure}
\begin{figure}
\includegraphics[width=\linewidth]{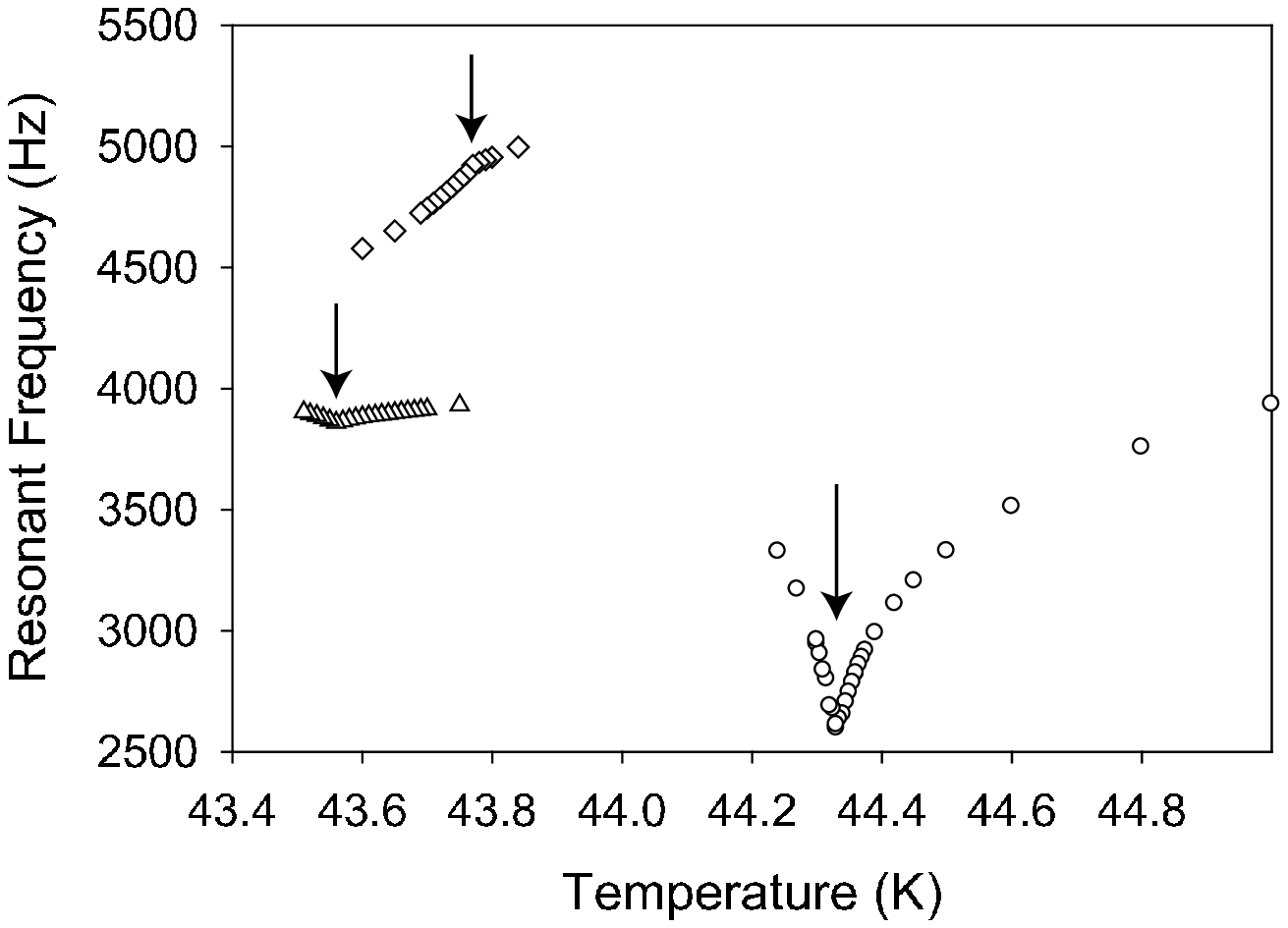}
\caption{\label{BulkNeonIsochores}Bulk neon isochores for
\mbox{$\rho = 282 \frac{kg}{m^3}$ (triangles),} \mbox{$\rho = 493
\frac{kg}{m^3}$ (circles),} and \mbox{$\rho = 655 \frac{kg}{m^3}$
(diamonds).}  Along each isochore the transition temperature is
indicated by a vertical arrow.}
\end{figure}

Figure~\ref{BulkNeonSpectrum} is an example of a spectrum, taken
with bulk neon near its critical density ($\rho_c \approx 485
\frac{kg}{m^3}$) at a temperature just above the liquid-vapor
critical point; the fundamental and first three harmonics are
labelled.  At higher frequencies there was interference from other
modes; only the low frequency resonances were used to track the
system's behavior. The frequency of the fundamental resonance was
recorded at different temperatures and these data were combined to
form an isochore.  As can be seen in
Figure~\ref{BulkNeonIsochores},
the shape of such an isochore depends on the density of the fluid.
The phase transition appeared as a very sharp dip when
$\rho\sim\rho_c$, but far from $\rho_c$ the transition appeared as
a less pronounced kink.

The position of the resonant peaks can be used to calculate the
approximate speed of sound in bulk neon as a function of
temperature along an isochore.  The fundamental and first harmonic
should give identical values for sound speed in a cylindrical
resonator, although there will be a slight offset due to the
non-ideality of our resonator. In our case the two peaks gave very
similar values for sound speed in the one phase region, but the
apparent ``sound speeds'' differ in the two phase region. This was
because the existence of an inhomogeneous fluid distribution, with
the appearance of a liquid-vapor meniscus, destroyed the
cylindrical symmetry of the resonator and the second resonant
frequency was no longer a simple harmonic of the fundamental
frequency. Apparent sound speed \mbox{(i.e.\ $\frac{2Lf}{n}$)}
along an isochore is plotted for $\rho_{neon} =670 \frac{kg}{m^3}$
in
\begin{figure}
\includegraphics[width=\linewidth]{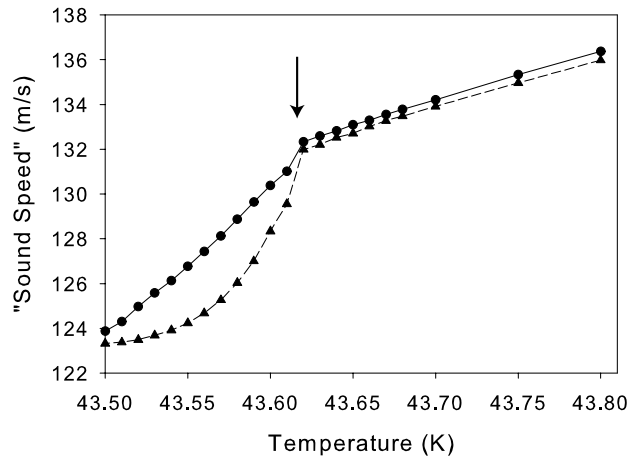}
\caption{\label{BulkNeonResSpeed}Apparent sound speed in bulk neon
along an isochore ($\rho = 670 \frac{kg}{m^3}$).  The values
calculated from the fundamental resonance (circles) and first
harmonic (triangles) are shown.  The liquid-vapor transition is
indicated by the vertical arrow.}
\end{figure}
Figure~\ref{BulkNeonResSpeed} --- below 43.620K the sound speeds
changed dramatically, marking the entrance into two-phase
coexistence.  Even when the resonant frequency did not show a
sharp feature at the transition, comparing apparent sound speeds
calculated from the first two resonant frequencies allowed us to
determine of the transition temperatures.

\subsection{Fluids in Aerogels}
Resonator results for the aerogel samples are of the same form as
the bulk neon results discussed above, but with complications. The
data points along an isochore were necessarily sparse because of
long equilibration times. Aerogel is an excellent thermal
insulator, which means experiments must use very small samples or
very slow data collection. While some
experiments~\cite{Wong90-2567,Herman05a} work well with thin
samples, our resonance experiments used large aerogels with their
associated long equilibration times. At temperatures above the
two-phase region, equilibration times of half an hour or less were
sufficient to reach equilibrium.  It was often impossible to reach
equilibrium within the two-phase region on the laboratory time
scale. This limited the temperature resolution --- with only one
or two dozen points along an isochore, the transition temperature
could not be determined to better than $\pm5$mK.

The addition of aerogel to the system also affected the form of
the resonance spectra.  With the aerogel backbone to help support
acoustic modes the transition did not appear as distinct as in the
pure fluid system; at low fluid densities the transition was often
undetectable. Another major complication added by the aerogel was
the appearance of peak splitting in many of the spectra.  There
appeared to be frequency windows in which the resonator could not
support a fluid resonance. In later experiments it was not
uncommon to encounter many such windows, and the frequency windows
shifted with fluid density.  These are discussed in more detail in
the section on helium in aerogel B51, where they were the most
problematic.

\subsection{Neon in Aerogel B110}
\begin{figure}
\includegraphics[width=\linewidth]{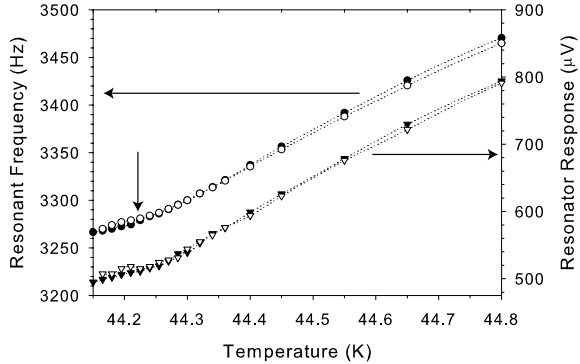}
\caption{\label{Neon95IsochoreLowD}Isochore of neon in aerogel
B110, $\rho_{Ne} = 472 \frac{kg}{m^3}$, showing the position and
amplitude of the lowest frequency resonance peak. The horizontal
arrows link the frequency (circles) and amplitude (triangles) data
to their respective axes.  The vertical arrow denotes the
transition temperature.  Data taken during cooling (solid symbols)
and warming (open symbols) are shown.}
\end{figure}

\begin{figure}
\includegraphics[width=\linewidth]{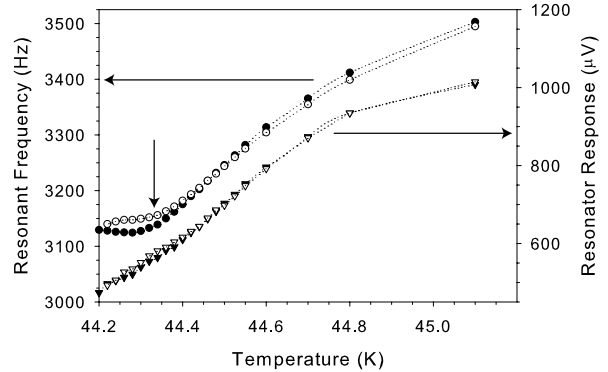}
\caption{\label{Neon95IsochoreMidD}Isochore of neon in aerogel
B110, $\rho_{Ne} = 535 \frac{kg}{m^3}$, near the center of its
``coexistence curve.'' Resonant frequency (circles) and amplitude
(triangles) are shown for cooling (solid symbols) and warming
(open symbols) runs.}
\end{figure}

\begin{figure}
\includegraphics[width=\linewidth]{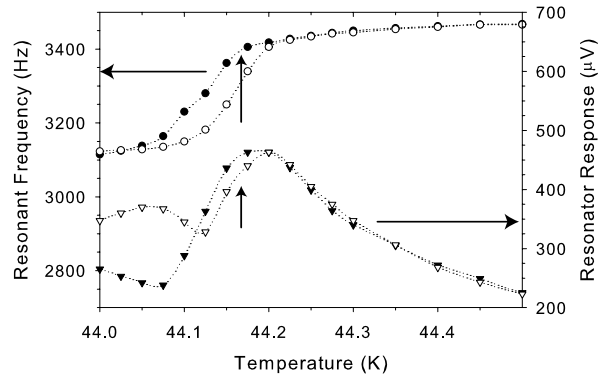}
\caption{\label{Neon95IsochoreHighD}Isochore of neon in aerogel
B110, $\rho_{Ne} = 641 \frac{kg}{m^3}$.  Resonant frequency
(circles) and amplitude (triangles) are shown for cooling (solid
symbols) and warming (open symbols) runs.}
\end{figure}

Some of our preliminary data\cite{Herman02} have been published
for neon in aerogel.  However, that aerogel had a density of
roughly $130\pm10 \frac{kg}{m^3}$, and was surrounded by a
significant ($\sim20\%$) bulk volume. The data presented here for
neon in aerogel B110 supersede those earlier data and represent
our highest precision resonator results.  Some representative
isochores are included as
Figures~\ref{Neon95IsochoreLowD}-\ref{Neon95IsochoreHighD}.
Each figure includes the frequency and amplitude of the acoustic
resonance as read from the \mbox{lock-in} amplifier. Absolute
values of the amplitude have little meaning, but relative
amplitudes along a single isochore give information on attenuation
of the acoustic signal. At the transition one expects a sudden
change in attenuation.  The highest fluid density used in this
experiment corresponded to the highest filling pressure that we
were certain the cell could withstand without damage. The lower
bound for the density was set by our ability to distinguish a
transition along the isochores.

From these three representative curves one can see that no
universal distinctive feature marks the liquid-vapor transition,
but that signs of a transition, indicated in
Figures~\ref{Neon95IsochoreLowD}-\ref{Neon95IsochoreHighD} by
vertical arrows, become more obvious as the density increases. At
low neon densities (for example, Fig.~\ref{Neon95IsochoreLowD}) it
is difficult to discern a feature which represents a phase
transition since even hysteresis between cooling and warming runs
falls below the resolution of the acoustic resonator, although the
slope in frequency does change at some point. By increasing the
density (shown in Fig.~\ref{Neon95IsochoreMidD}) the change of
slope is more obvious, and hysteresis becomes visible. Finally, at
high densities (e.g.\ Fig.~\ref{Neon95IsochoreHighD}) there is a
significant, and sharp, change in resonant frequency at the
transition.  Hysteresis in these isochores indicates
non-equilibrium behavior --- below the liquid-vapor transition
equilibration times stretched into hours so it is unclear whether
this hysteresis would disappear given enough time or if it
represents rate-independent behavior.

For each feature in the resonant frequency there should be a
corresponding feature in the peak amplitude. At the transition one
expects a sharp increase in attenuation, and consequently a sharp
decrease in amplitude. In the lower density curves
(Fig.~\ref{Neon95IsochoreLowD} and Fig.~\ref{Neon95IsochoreMidD}),
such a sharp decrease is not seen, although in both cases the
amplitude curves reveal slight indications of a transition at the
same temperature as seen in the frequency curves (i.e. change of
slope or appearance of hysteresis in the peak amplitude).  In the
high density curve (Fig.~\ref{Neon95IsochoreHighD}) however, there
is a sharp decrease in amplitude at the same temperature as the
sharp drop in resonant frequency, although it is somewhat obscured
because the resonance is split between the peak near 3450Hz and
another at 3600Hz (not shown).  This peak splitting behavior is
described in more detail later, for helium in B51.

Both cooling and warming runs are plotted in the figures, and both
show the liquid-vapor transition, but often at slightly different
temperatures. Since the warming runs started in the two phase
region where latent heat may have been released or absorbed as the
temperature was changed, thermal equilibrium was not established
as quickly as in the single phase system. When picking the
transition temperature for each isochore the precise choice was
always made using the cooling data.

\begin{figure}
\includegraphics[width=\linewidth]{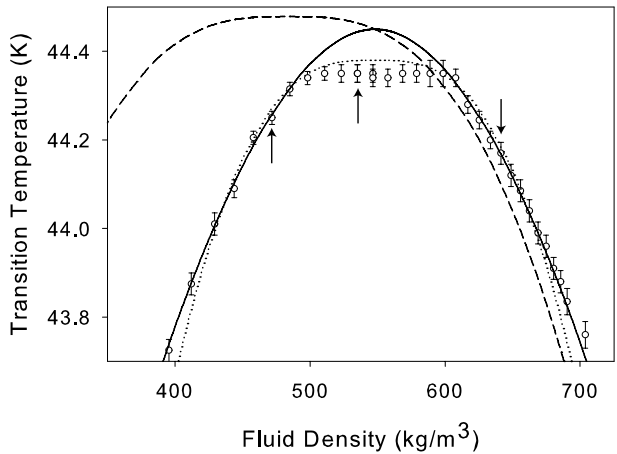}
\caption{\label{Neon95ResCCurve}Neon coexistence curve in aerogel
B110.  These raw data (circles) do not take into account the error
in the thermometer calibration.  Two fits are shown, assuming
$\beta = 0.5$ (solid line) and $\beta = 0.325$ (dotted line). The
dashed line is the accepted coexistence curve of bulk
neon\cite{Pestak84-274}.  The three points corresponding to the
isochores in
Figures~\ref{Neon95IsochoreLowD}-\ref{Neon95IsochoreHighD} are
indicated by the vertical arrows.}
\end{figure}

The final ``coexistence curve'' for neon in aerogel B110 is shown
in Figure~\ref{Neon95ResCCurve}; the three points corresponding to
the isochores shown in
\mbox{Figures~\ref{Neon95IsochoreLowD}-\ref{Neon95IsochoreHighD}}
are indicated by vertical arrows.
There are three important differences between the bulk neon
coexistence curve (shown by the dashed line) and the data for neon
in aerogel.  The first is that the curve has been narrowed, the
second is that the center of the curve has been shifted to higher
density, and the third is that the shape of the the coexistence
region seems to include a flat region in the center.

There is also a shift in transition temperature, although with the
uncertainty about the calibration of the thermometer it difficult
to say much about that shift other than observing that it is
small. Figure~\ref{Neon95ResCCurve} includes the raw data, with no
adjustments made to the thermometer calibration.  Our calibration
appeared to be off by more than 100mK, so if the adjusted
temperatures are used then the coexistence curve has only been
depressed by 0.02K, or 0.05\%. This very small shift is an order
of magnitude smaller than the shift seen for nitrogen (0.6\%) in
Ref.~\cite{ApolloWongPhd,Wong93}.

Without a fit to the coexistence curve a quantitative analysis of
its width is not possible, however, a cursory examination of the
curve indicates only slight narrowing with respect to bulk
\mbox{($<50\%$)}. This narrowing is significantly less than that
reported in helium but only slightly less than
nitrogen\cite{ApolloWongPhd,Wong93}. The plateau seen at the top
of the curve is also a new feature, but may reflect the changing
nature of the transition marker rather than a fundamental change
in behavior. If we do not include the data along the plateau for
now, and just focus on the low and high density sides of the
coexistence curve for neon in aerogel, we can evaluate the
critical exponent $\beta$. This would be equal to $\sim0.325$ for
a bulk system close to its critical point or 0.5 for a mean field
transition; fits with $\beta=0.325$ and $\beta=0.5$ are included
in the figure.

The fit assuming Ising behavior (i.e.\ $\beta\sim0.33$) intersects
most of the data within error, but not all. The second fit, with
\mbox{$\beta=0.5$}, accommodates the data along the low and high
density sides of the curve much better but does not intersect data
near the center of the curve.  Since the existence of a true
equilibrium liquid-vapor transition for fluids in aerogel is still
an open question, the power-law fits in
Figure~\ref{Neon95ResCCurve} are only shown to help describe the
shape of the curve, not to imply a certain type of critical
behavior.

\subsection{Helium in Aerogel B110}
\begin{figure}
\includegraphics[width=\linewidth]{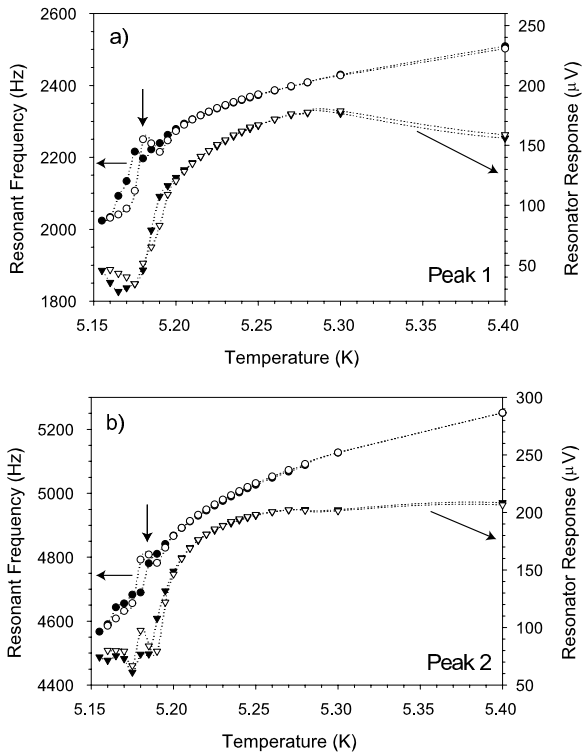}
\caption{\label{Helium95IsochoreMidD}Isochore of helium in aerogel
B110, $\rho_{He} = 88.6 \frac{kg}{m^3}$ near the center of the
``coexistence curve.'' Data are shown for the (a) lowest frequency
resonance, and (b) the second resonance.  Cooling (solid symbols)
and warming (open symbols) data are shown for resonant frequency
(circles) and amplitude (triangles). There are indications of the
phase transition in the resonant frequency and peak amplitude of
both peaks 1 and 2.}
\end{figure}

Isochores were taken with helium in a resonator filled with a
second sample of aerogel B110; a typical isochore is included as
Figure~\ref{Helium95IsochoreMidD}.
Again arrows link data to the corresponding axis and vertical
arrows indicate the transition. Whenever the first two resonances
were observable, both were analyzed; by taking advantage of this
redundancy it was possible to pick out transition temperatures
even if the transition was obscured along one of the resonances.
Most points within this data set were collected after a 15 minute
equilibration time --- that was sufficient time for points above
the transition to equilibrate; points below the transition did not
always have time to equilibrate, leading to hysteresis at lower
temperatures.

\begin{figure}
\includegraphics[width=\linewidth]{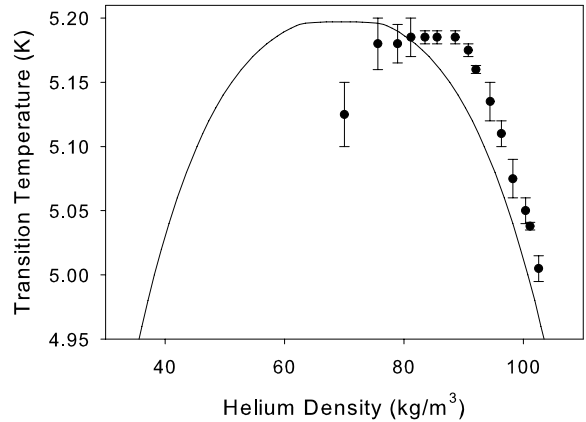}
\caption{\label{Helium95ResCCurve}Coexistence curve of helium in
aerogel B110. The bulk coexistence curve is included as a solid
line~\cite{NISTWebBook}.}
\end{figure}

Isochores were used to construct the coexistence curve shown in
Figure~\ref{Helium95ResCCurve}.  The error bars reflect both the
uncertainty in picking a transition temperature from a rounded
feature and any differences in the transition temperature
determined from the first and second resonant peaks.
The curve is narrowed by about a factor of about two from bulk and
it is shifted toward the higher density side of the bulk helium
coexistence curve. The data are too uncertain to make meaningful
fits to critical exponents, but are not inconsistent with an
Ising-like fit. They are also not precise enough to state
confidently whether there is a plateau at the center of the curve
as seen in the neon data.

\subsection{Helium in Aerogel B51}
\begin{figure}
\includegraphics[width=\linewidth]{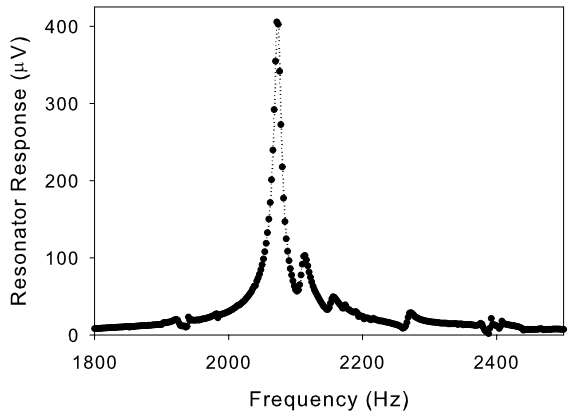}
\caption{\label{Helium98Spectrum}Spectrum of helium ($\rho_{He} =
93.1 \frac{kg}{m^3}$) in aerogel B51, at 5.140K, showing
additional resonance peaks resulting from the interference of
acoustic modes in the fluid-filled aerogel.}
\end{figure}

The results for the resonator experiments on helium in aerogel B51
resemble those for helium in aerogel B110.  However, the resonant
peaks were more difficult to track, with more interference;
Fig.~\ref{Helium98Spectrum} shows an example of a single spectrum.
While there is a single large peak at $\sim 2080$Hz, there are
also peaks at higher frequencies that are a result of interference
with very low frequency modes. Rather than a single resonant
frequency that moved to lower frequencies as the sample was
cooled, there would usually be a number of peaks and the relative
amplitudes of these peaks would vary with temperature.  The
resonant frequencies were generally lower than in B110, because of
the lower elastic constants of the less dense aerogel, and any
signal above about 5kHz was obscured by interference between lower
frequency resonances.

The ubiquity of the interference in this data set provides an
excellent opportunity to analyze that behavior and its possible
origin. There were often frequency windows in which the B51
resonator could not support a longitudinal resonance, consistent
with mode crossing.  There were problems with interference in both
the neon and helium data sets in B110, however in neither case was
the problem as severe.

The interfering modes were probably not external resonances (such
as membrane resonances) since they shifted monotonically with
fluid density, but exactly what internal modes were interfering is
not clear.
\begin{figure}
\includegraphics[width=\linewidth]{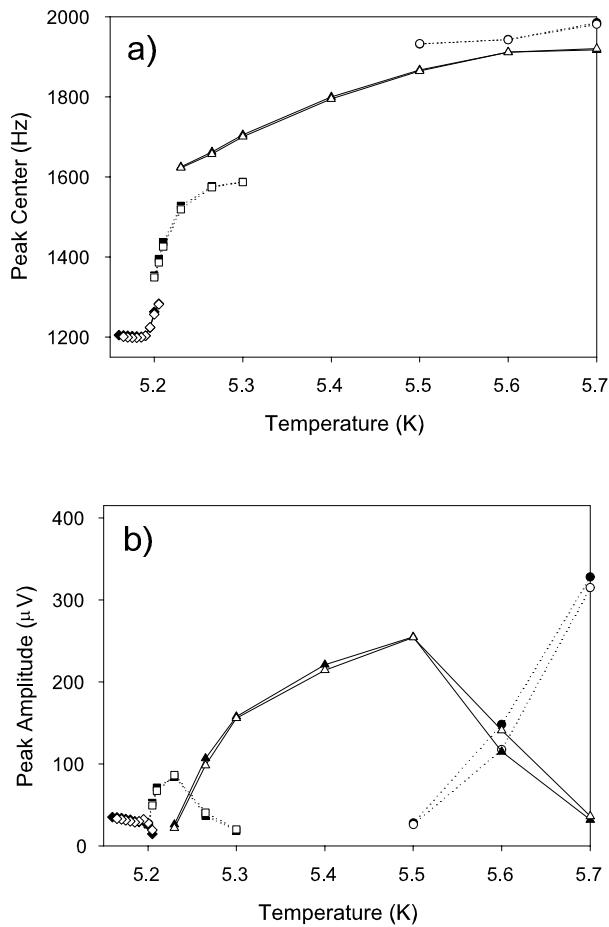}
\caption{\label{Helium98IsochoreLowD1}Isochore of helium in
aerogel B51, $\rho_{He} = 73.1 \frac{kg}{m^3}$, showing multiple
mode crossings. The two panels show (a) frequency and (b)
amplitude of the resonant peaks during cooling (solid symbols) and
warming (open symbols).}
\end{figure}
In Figure~\ref{Helium98IsochoreLowD1} a complete isochore of
helium in aerogel B51 is shown to emphasize the effects of these
interfering resonances.  In Fig.~\ref{Helium98IsochoreLowD1}a
there are four distinct peak regions; as the longitudinal acoustic
resonance shifts to lower frequencies (because the compressibility
of the fluid is increasing as temperature falls) it encounters
three frequency regions in which no peak exists. These regions,
near 1300Hz, 1600Hz, and 1925Hz respectively, probably represent
frequencies at which mode crossing occurred.  If so, the
dependence of these ``forbidden'' frequencies on density should
give a hint as to the form of the other resonance present in the
cell.

\begin{figure}
\includegraphics[width=\linewidth]{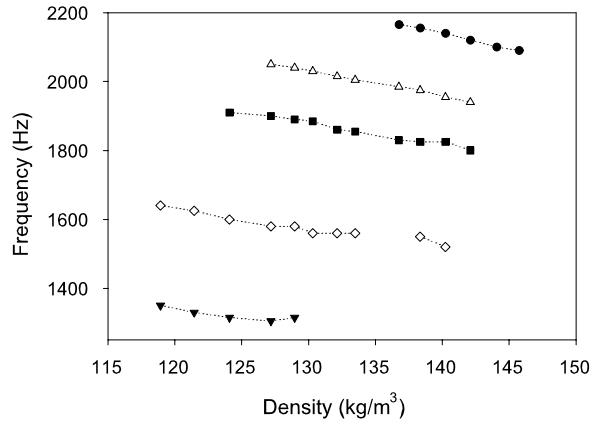}
\caption{\label{Helium98ModeCrossing}Mode crossing in aerogel B51
isochores.  Here the forbidden frequencies are plotted as a
function of total system density ($\rho_{gel} + \rho_{He}$).}
\end{figure}

The spacing between modes ($\sim300Hz$) implies that they are
governed by very low elastic constants; since we are already
probing the lowest longitudinal acoustic mode of the
aerogel-filled cavity the obvious candidates are shear modes.
Using values of the shear modulus of aerogel\cite{Daughton03}, a
simple torsional oscillation of a solid cylinder with dimensions
similar to our aerogels should have a resonance near 1000Hz,
larger than we saw but of the same order of magnitude. If we
assume that the modes responsible are mostly shear in nature, then
the restoring force is unaffected by the density of fluid in the
cell and the frequency should be inversely proportional to the
root of the system density ($f \propto \sqrt{\frac{\mu}{\rho}}
\propto \rho^{- \frac{1}{2}} $). In
Figure~\ref{Helium98ModeCrossing} the frequencies of mode
crossings are plotted as a function of system density (i.e.\ the
sum of the densities of the fluid and the aerogel).

The density dependence of the interfering modes is consistent with
them being mostly shear in character, although with the current
data set there is no way to state with confidence what the form of
these resonances actually was.  The large number of modes wreaks
havoc with choosing transition temperatures from the isochores;
while in the isochores in aerogel B110 there were a couple of
regions of density where the transition was obscured, in aerogel
B51 every isochore shows at least some evidence of mode crossing
near the transition.

\begin{figure}
\includegraphics[width=\linewidth]{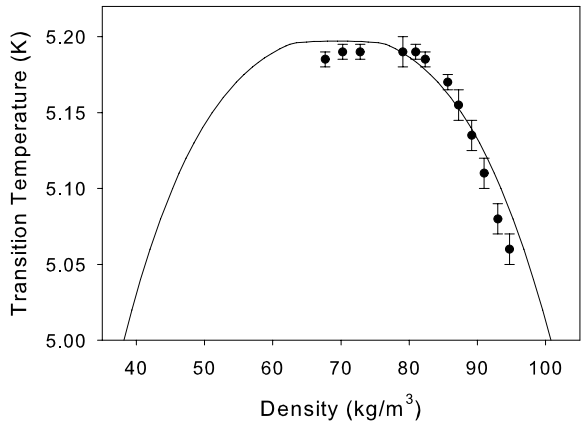}
\caption{\label{Helium98ResCCurve}``Coexistence curve'' of helium
in aerogel B51.}
\end{figure}

Other than the increased interference, the form of the isochores
was the same as in aerogel B110 and no single feature presented
itself as an obvious marker of a transition. Some isochores showed
obvious hysteresis, due in part to slow equilibration below the
transition --- in such cases the cooling data were taken as being
closest to equilibrium near the transition. Since the form of the
isochores has already been shown, no additional isochores are
shown for this sample, but the ``coexistence curve'' for helium in
aerogel B51 extracted from the resonator isochores is shown in
Figure~\ref{Helium98ResCCurve}.
 The lack of low density points along the curve is
due to the lack of distinct markers for any sort of transition
along the low density isochores. Subtle changes in slope of the
isochore which could be used in the B110 resonator isochores were
swamped by interference or mode crossing in the B51 resonator. The
curve is only well defined on the high density side, although the
lowest density point may indicate that the transition temperature
is starting to move to lower values.  This would imply that the
low density branch of the curve was shifted significantly from the
bulk curve, but not as much as in helium in aerogel B110;
certainly the curve is wider than in B110. The lack of a clear
marker for the transition on the low density side is consistent
with results from the other resonators, and with the gradual
transition seen on the low density side of adsorption
isotherms\cite{Herman05a}. No attempts to fit the curve to
critical exponents have been made.

\section{Conclusions/Discussion}

In this study a set of low frequency acoustic experiments were
undertaken to study the liquid-vapor transition of fluids near
their critical points in silica aerogel.  This technique is
capable of pinpointing the onset of liquid-vapor coexistence along
isochores in bulk fluids but the presence of an aerogel
complicated the resonance modes and made the transition less
obvious.  The transitions were not as sharp as for bulk fluids,
particularly on the low density ``vapor'' branch.  Phase
separation was accompanied by hysteresis and long time constants,
behavior consistent with that seen in adsorption isotherm studies
of fluids in aerogels.  Nonetheless, by following isochores and
looking at the dissipation and the frequencies of fundamental and
higher resonance modes, we were able to find markers of phase
separation and map out ``coexistence curves'' for neon and for
helium in two different aerogels.  Ambiguities about which
features to include in a ``coexistence curve'' are common to most
studies of liquid-vapor behavior in porous media but such curves
allow a quick comparison to other experiments and to bulk fluid
behavior. However, it is important to remember that curves based
on the appearance of hysteresis or other transition markers may
not represent a macroscopic phase transition with equilibrium
coexistence between liquid and vapor-like phases.  We will
therefore only make qualitative comparisons to the results of
other experiments.

Previous measurements\cite{Wong90-2567,Wong93,Gabay00-585} of
liquid-vapor behavior have been made using aerogels with similar
density to our sample B110 ($110 \frac{kg}{m3}$, 95\% porosity).
Gabay \textit{et al}.~\cite{ClaudeGabayPhd,Gabay00-585} used a
mechanical oscillator to measure the helium density while
isothermally flowing helium at different rates into or out of
their aerogel.  They observed hysteresis even quite close to the
LVCP and found no evidence for an equilibrium liquid-vapor
transition. Wong and Chan, on the other hand, saw qualitatively
different behavior~\cite{Wong90-2567}.  They measured heat
capacity along isochores, following similar thermodynamic paths to
the resonance measurements in this paper, and found an extremely
narrow coexistence curve near the LVCP, with the liquid branch
shifted to lower density than bulk liquid.  They supplemented
these measurements with several adsorption isotherms which were
consistent with liquid-vapor coexistence, and reported no
hysteresis.

Our measurements in aerogel B110 are consistent with the behavior
seen by Gabay et al. and with our own isotherm
measurements\cite{Herman05a}.  We see hysteresis below the
liquid-vapor transition and a coexistence curve which is somewhat
narrower than for bulk fluids and shifted to higher densities.  As
one might expect, the higher porosity aerogel B51 has a less
pronounced effect, with a high density liquid branch which is
quite close to that in bulk helium.  Since the transition markers
were very subtle at low densities, it was not possible to map out
the vapor branch over a sufficiently wide range to make
quantitative statements about the widths of the coexistence
regions for helium. However, we do not see any evidence for the
extremely narrow coexistence curve reported in
Ref.~\cite{Wong90-2567}, even though we used similar samples and
thermodynamic paths.  It is possible that the long thermal time
constants we observed in these and our earlier isotherm studies
affected the heat capacity results.  In AC heat capacity
measurements only a thin layer of helium near the surface of the
sample would respond, even at a heater frequency as low as 0.1 Hz,
in contrast to acoustic resonances where adiabatic compression
probes the entire sample.

Mean field simulations of liquid-vapor behavior in porous media
show a ``hysteresis phase diagram'' (analogous to a coexistence
curve) with a width that depends on the ``wettability'' $y =
\frac{w_{fs}}{w_{ff}}$, the ratio between the strengths of the
adsorbate-substrate (fluid-solid) and adsorbate-adsorbate
(fluid-fluid) attractive interactions, $w_{fs}$ and $w_{ff}$.  As
\textit{y} increases, the coexistence region becomes narrower.
Although the simulations are coarse-grained, so \textit{y} cannot
be computed directly from van der Waals interactions, it can still
be varied by choosing different fluids.  One measure of the
adsorbate-substrate attraction $w_{fs}$ is the coefficient
$\gamma$ of the long-range d$^{-3}$ van der Waals potential
between a fluid atom and a glass surface (1100 and 1640 K\AA $^3$
for helium and neon respectively\cite{Cheng88}).  The fluid-fluid
interaction is reflected in properties such as surface tension
(0.3 and 6 $\frac{mN}{m}$  for helium and neon respectively at low
temperature\cite{NISTWebBook}) and the latent heat of
vaporization, $L_v$ (2.6 and 103 $\frac{J}{cm^3}$ for helium and
neon respectively).  If we characterize fluids by the ratio
$\frac{\gamma}{L_v}$, we expect helium to have a much larger
wettability than neon ($\frac{\gamma}{L_v}$ about 27 times
larger).  Nitrogen, for comparison, has a ratio
$\frac{\gamma}{L_v}$ close to that of neon.  Our neon coexistence
curve in Fig.~\ref{Neon95ResCCurve} does have a width similar to
that seen by Wong et al. in their light scattering
experiments\cite{Wong93} with nitrogen.  However, although the
helium coexistence curve in Fig.~\ref{Helium95ResCCurve} appears
to be narrower than that for neon, the difference is not as large
as would be expected from the simulations given the large
difference between the wettability of neon and helium.

The neon measurements presented in this paper are the most
detailed and precise that we have made.  However, it is not clear
that the transitions we saw correspond to true liquid-vapor
equilibrium, so we have not attempted to extract critical
exponents from the data near the LVCP.  As
Fig.~\ref{Neon95ResCCurve} shows, the shape of the curve is
roughly consistent with an Ising-like fit over the entire density
range, but could also be fit with a mean-field exponent except for
the flat plateau region near the critical density.

Our acoustic resonance experiments provide information about
liquid-vapor behavior in aerogels, to complement that available
from recent adsorption isotherm measurements. The behavior is
consistent with the isotherm results, with a relatively sharp
transition on the high density side of the ``coexistence curves''
but more gradual changes on the low density side. This, together
with the hysteresis we usually observed below the transitions,
makes it impossible to unambiguously define a coexistence curve or
even to be sure we can access the equilibrium states that would be
required for a true equilibrium phase transition. The observed
behavior was similar for helium and neon, despite their different
interaction strengths. The lower density aerogel had a smaller
effect on helium's coexistence curve, but the general features of
the transition were the same. Despite their tenuous nature, it
does not appear that aerogels behave as a simple dilute impurity
which can be used to study the equilibrium liquid-vapor critical
behavior in the presence of disorder. The observed hysteresis
indicates the importance of metastable states and suggests that a
description in terms of capillary condensation is more
appropriate.

\end{document}